\documentclass[showpacs,epsf,twocolumn]{revtex4-1}
\usepackage{float}
\usepackage{color,graphicx}
\graphicspath{ {images/} }
\usepackage{hyperref}
\usepackage{xcolor}
\usepackage{framed}
\colorlet{shadecolor}{blue!20}
\begin{document}

\title{\textbf{Large nonsaturating magnetoresistance and signature of non-degenerate Dirac nodes in ZrSiS}}

\author{\textbf{Ratnadwip Singha$^{a}$, Arnab Pariari$^{a}$, Biswarup Satpati$^{a}$, Prabhat Mandal$^{a}$}}

\affiliation{$^{a}$Saha Institute of Nuclear Physics, HBNI, 1/AF Bidhannagar, Kolkata 700 064, India.}

\maketitle

\textbf{While the discovery of Dirac and Weyl type excitations in electronic systems is a major breakthrough in recent condensed matter physics, finding appropriate materials for fundamental physics and technological applications, is an experimental challenge. In all the reported materials, linear dispersion survives only up to a few hundred meV from the Dirac or Weyl nodes. On the other hand, real materials are subject to uncontrolled doping during preparation and thermal effect near room temperature can hinder the rich physics. In ZrSiS, ARPES measurements have shown an unusually robust linear dispersion (up to $\sim$2 eV) with multiple non-degenerate Dirac nodes. In this context, we present the magnetotransport study on ZrSiS crystal, which represents a large family of materials (\textit{WHM} with \textit{W} = Zr, Hf; \textit{H} = Si, Ge, Sn; \textit{M} = O, S, Se, Te) with identical band topology. Along with extremely large and non-saturating magnetoresistance (MR), $\sim$ 1.4 $\times$ 10$^{5}$ \% at 2 K and 9 T, it shows strong anisotropy depending on the direction of the magnetic field. Quantum oscillation and Hall effect measurements have revealed large hole and small electron Fermi pockets. Non-trivial $\pi$ Berry phase confirms the Dirac fermionic nature for both types of charge carriers. The long-sought relativistic phenomenon of massless Dirac fermions, known as Adler-Bell-Jackiw chiral anomaly, has also been observed.}

The discovery of topological insulators (1) and three-dimensional Dirac and Weyl semimetals (2,3), has emerged as one of the major breakthroughs in condensed matter physics in recent time. Materials with topologically non-trivial band structure serve as template to explore the quantum dynamics of relativistic particles in low-energy condensed matter systems. In addition to rich physics, these systems offer possibility of practical applications in magnetic memory, magnetic sensor or switch and spintronics, due to the novel transport phenomena such as extreme magnetoresistance and ultrahigh mobility (4-6). In Dirac semimetals, bulk valence and conduction band undergo linear band crossings at four-fold degenerate Dirac points protected by time reversal (TRS), inversion (IS) and crystal symmetry (CS) (7,8). By breaking either TRS or IS, each Dirac point can be broken into a pair of doubly degenerate Weyl points, accompanied by the surface Fermi arc (7,8). Theoretical prediction (7,8) followed by ARPES and transport measurements have verified the existence of bulk Dirac points in Cd$_{3}$As$_{2}$ and Na$_{3}$Bi (2, 9-11) and Weyl nodes in IS breaking $TX$ ($T$ = Ta, Nb; $X$ = As, P) family of materials (3,12-14) and TRS breaking YbMnBi$_{2}$ (15). Apart from these compounds, recently, topological nodal line semimetals (TNLSM) have emerged, where the bands cross along one-dimensional closed lines in \textbf{k}-space instead of discrete points. Although proposed in few materials (16,17), the existence of nodal line has been experimentally verified only in IS breaking non-centrosymmetric superconductor PbTaSe$_{2}$ (18).

Recently, first-principle calculations and ARPES measurement have revealed the existence of multiple Dirac crossings along with unconventional hybridization of surface and bulk states in ZrSiS (19,20). The Dirac nodes are protected by non-symmorphic symmetry and reside at different energy values of band structure with a diamond shaped Fermi surface. Another feature which makes ZrSiS an interesting system, is the energy range of the linear band dispersion. While most of the materials observed so far have linear band dispersion up to a few hundred meV from the Dirac point, in ZrSiS the range is observed to be as high as 2 eV in some regions of the Brillouin zone. To realize and exploit the interesting features of Dirac or Weyl fermions in electronic transport properties, the primary requirement is that the Fermi energy of the material should remain within the linear dispersion region. As real materials often undergo uncontrolled doping or deviation from ideal stoichiometry during preparation, very careful and delicate experimental procedures are required to ensure that this primary criterion is fulfilled. On the other hand, very large energy range of linear band dispersion makes ZrSiS robust enough to satisfy this requirement even when the crystals encounter certain level of carrier doping or non-stoichiometry. Hence, ZrSiS represents a sturdy topological system, which can be used in industrial applications.

\textbf{Results}

\textbf{Sample characterization.} High-resolution transmission electron microscopy (HRTEM) and energy-dispersive X-ray spectroscopy (EDX) reveal high quality of the ZrSiS single crystals without any impurity. The details are given in SI (Fig. S1 and Fig. S2).

\textbf{Temperature dependence of resistivity.} As shown in Fig. 1\textit{A}, the zero-field resistivity of ZrSiS shows metallic character. $\rho$ decreases monotonically with the decrease in \textit{T} down to 2 K. The resistivity at 2 K becomes as low as $\sim$ 52 n$\Omega$ cm which is comparable to that reported for Cd$_{3}$As$_{2}$ (10). At temperature below 10 K, the measured resistivity shows some fluctuations within the instrument resolution, which can be explained in terms of the quantum ballistic transport (21). The ultralow residual resistivity and signature of quantum ballistic transport suggest that the mean free path of the charge carriers is very large. Hence, the impurity effect in ZrSiS is almost negligible. The residual resistivity ratio $\rho$(300 K)/$\rho$(2 K) is found to be 288, which is quite large and confirms good metallicity and high quality of the crystals. The resistivity data, in the temperature range 10 to 115 K, can be fitted well with the expression $\rho(T) = \rho_{0} + AT^n$ with \textit{n} $\sim$ 3, as shown in Fig. S3\textit{A} (See SI). This indicates a deviation from pure electronic correlation dominated scattering mechanism (\textit{n} = 2) (22). Similar type of temperature dependence of $\rho$ has also been observed in unconventional semimetals LaSb (\textit{n} = 4) (23) and LaBi (\textit{n} = 3) (24) and has been attributed to interband electron-phonon scattering. $\rho$(\textit{T}) is linear in the high temperature region above 115 K. With the application of magnetic field, the low-temperature resistivity  undergoes a drastic enhancement, reflecting a metal-semiconductor-like crossover even at a field of 1 T only. This type of magnetic field-induced crossover is often described as a result of gap opening at the band touching points in topological semimetals (14, 23-26). It is evident from Fig. 1\textit{B} that the metal-semiconductor-like crossover is extremely sensitive to the direction of applied field. With current along \textbf{a} axis and magnetic field parallel to \textbf{c} axis, a strong crossover has been seen. On the other hand, rotating the field direction by 90$^{\circ}$, i.e., parallel to \textbf{b} axis, results in much weaker crossover, which occurs at higher field strength. In both the cases, the crossover temperature (\textit{T$_{m}$}) increases monotonically with field and is showing $T_{m} \propto (B - B_{0})^{1/\nu}$ type relation (Fig. S3\textit{B}) (See SI). $\nu$ has a value $\sim$3 for both the applied field directions and deviates from the value $\nu$ = 2 for compensated semimetals Bi, WTe$_{2}$ and graphite (27,28). Considering the thermal activated transport as in the case of intrinsic semiconductor (29), $\rho(T)$ = $\rho_{0} exp (E_{g}/2k_{B}T)$, we have calculated the values of the thermal activation energy gap $\sim$20.2 meV and $\sim$3.7 meV at 9 T, for field directions along \textbf{c} axis and \textbf{b} axis, respectively (Fig. S4\textit{A} and S4\textit{B}). The calculated gap \textit{E$_{g}$} shows strong magnetic field dependence (Fig. S4\textit{C}) (See SI). For both the directions, below \textit{T$_{m}$}, the resistivity exhibits an inflection followed by a plateau region. Similar low-temperature resistivity plateau has been observed in other topological semimetals and is independent of the sample quality (14, 23-26). Therefore, this low-temperature resistivity saturation is an intrinsic property of topological semimetals. However, the origin of this behavior is not yet settled (23).
\begin{figure}
\includegraphics[width=0.45\textwidth]{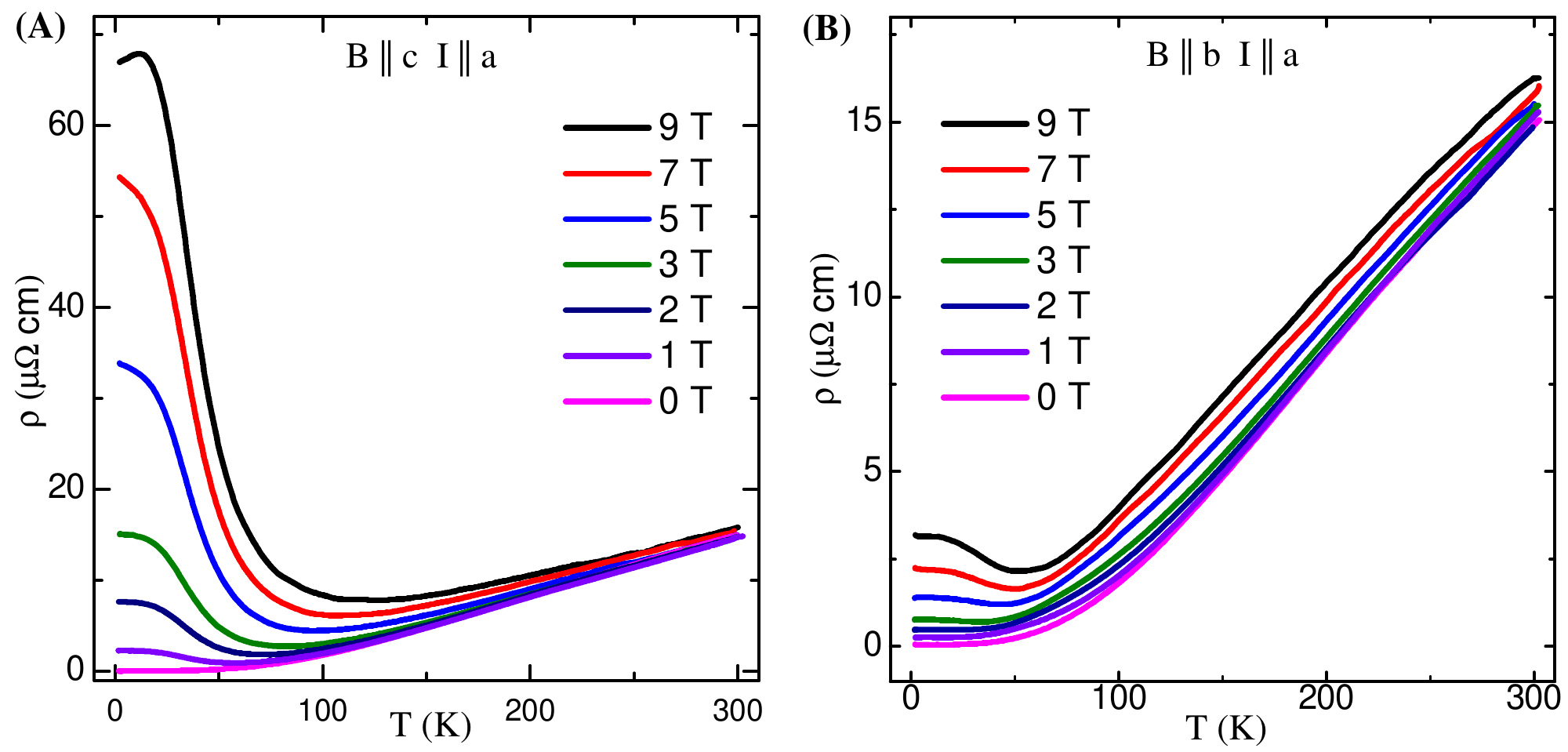}
\caption{Temperature dependence of resistivity measured under different transverse magnetic fields: \textit{(A)} \textit{\textbf{B}} $\parallel$ \textbf{c} axis and \textit{(B)} \textit{\textbf{B}} $\parallel$ \textbf{b} axis.}
\end{figure}

\textbf{Extreme transverse magnetoresistance.} The transverse magnetoresistance (TMR), i.e., the change in resistance with magnetic field applied perpendicular to the current direction has been measured at several temperatures. As illustrated in Fig. 2\textit{A}, at some representative temperatures, with current  parallel to \textbf{a} axis and magnetic field along \textbf{c} axis, an extremely large, non saturating MR is obtained. At 2 K and 9 T, MR is 1.4 $\times$ 10$^{5}$ \%, which is comparable to that observed in several Dirac and Weyl semimetals (10,13,14,25,26). With the increase in temperature, MR decreases dramatically to a value of just about 14 \% at 300 K and 9 T. At low field, MR shows a quadratic field dependence ($\propto$ B$^{2}$), which becomes almost linear at higher field. As shown in Fig. S4\textit{D} (See SI), the MR data at different temperatures can not be rescaled to a single curve using the Kohler's rule MR = $\alpha (\mu_{0}H/\rho_{0})^m$. The violation of Kohler's rule suggests the presence of more than one type of carrier and or the different temperature dependence of their mobilities (28,30). Applying field parallel to \textbf{b} axis and keeping the current direction unchanged, the MR at 9 T has been seen to reduce to $\sim$ 7000 \% at 5 K (Fig. 2\textit{B}). At 3 K and 9 T, the anisotropic ratio $\rho$(\textit{\textbf{B}} $\parallel$ \textbf{c})/$\rho$(\textit{\textbf{B}} $\parallel$ \textbf{b}) has a large value 21, which is comparable to that reported in NbSb$_{2}$ (26). This reflects strong anisotropy in electronic structure associated with the quasi two-dimensional nature of the Fermi surface observed in ARPES (20). For a two-dimensional system, where the charge is confined within the plane, the electronic motion is unaffected for magnetic field parallel to the plane, i.e., the anisotropy in MR will be extremely large. On the other hand, MR ratio is expected to be close to 1 for an isotropic three-dimensional system.

\begin{figure}
\includegraphics[width=0.45\textwidth]{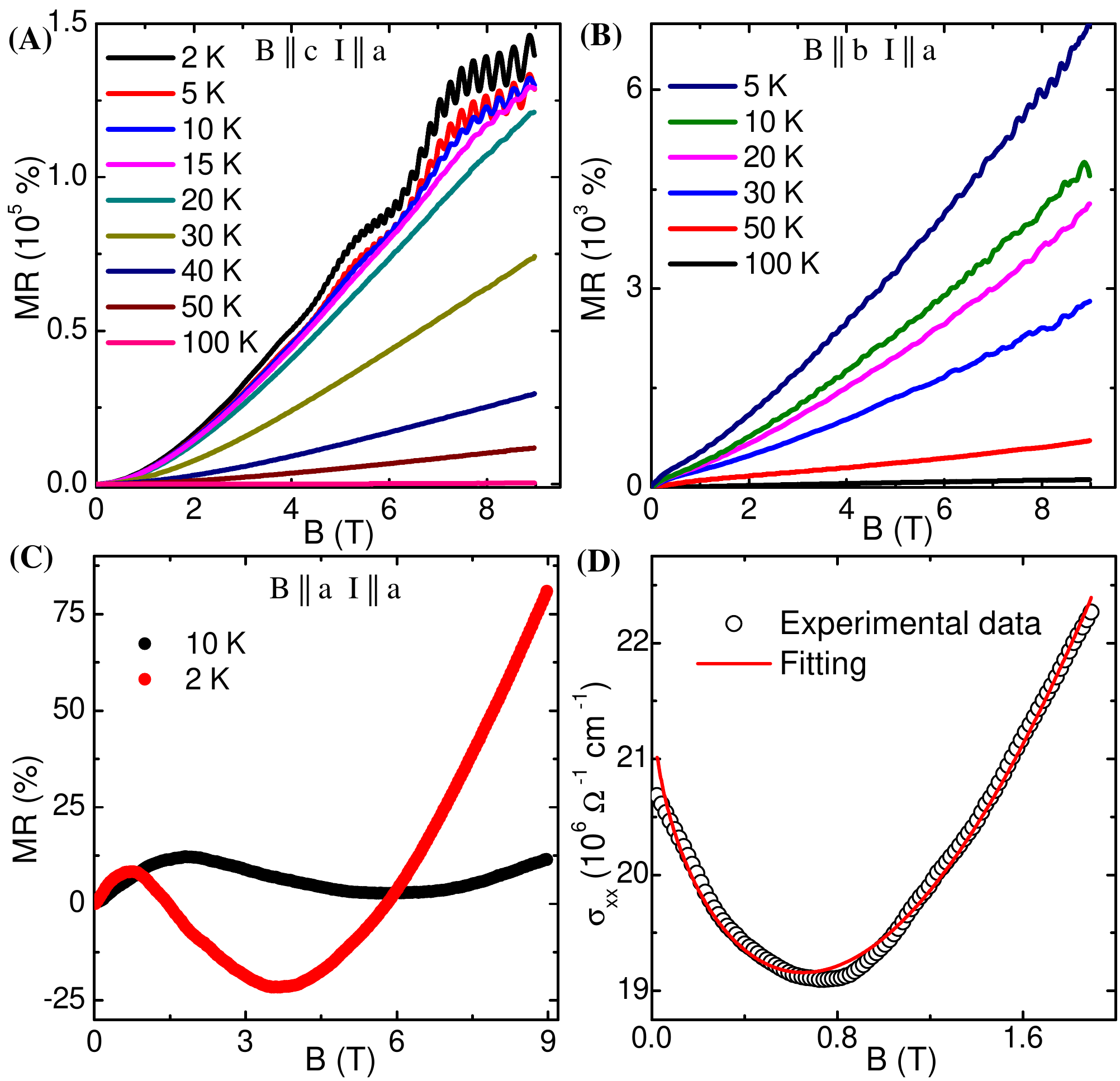}
\caption{Transverse magnetoresistance with current along \textbf{a} axis and magnetic field parallel to \textit{(A)} \textbf{c} axis and \textit{(B)} \textbf{b} axis, measured at different temperatures, up to 9 T. \textit{(C)} Longitudinal magnetoresistance (LMR) with current and field along \textbf{a} axis. \textit{(D)} Fitting of longitudinal magnetoconductivity data at 2 K using semiclassical formula.}
\end{figure}

\textbf{Longitudinal magnetoresistance and chiral anomaly.} Next, the longitudinal MR (LMR) has been measured with both the current and magnetic fields applied along \textbf{a} axis. As shown in Fig. 2\textit{C}, negative MR has been observed at low field. With the increase in temperature, the negative MR progressively weakens. This negative MR has been ascribed to induced Adler-Bell-Jackiw chiral anomaly in Dirac systems, where a Dirac node splits into two Weyl nodes with opposite chirality due to broken TRS, under application of magnetic field (8). Parallel magnetic (\textit{\textbf{B}}) and electric field (\textit{\textbf{E}}) act as a non-trivial gauge field (\textit{\textbf{E}.\textbf{B}}), which induces the chiral anomaly, i.e., charge imbalance between the two Weyl nodes of opposite chirality. This causes an extra flow of current along the direction of the applied electric field and results in the negative LMR. The chiral magnetic effect, a long sought phenomenon proposed in relativistic quantum field theory, has been demonstrated in several 3D Dirac and Weyl semimetals such as Cd$_{3}$As$_{2}$ (31), Na$_{3}$Bi (32) and TaAs (13). The field dependence of longitudinal conductivity [\textit{$\sigma$$_{xx}$(B$_{x}$)}] at a particular temperature, can be analyzed using the semiclassical formula (13),
\begin{equation}
\sigma_{xx}(B_{x}) = (1 + C_{w}B_{x}^{2})(\sigma_{0} + a\sqrt{B_{x}}) + (\rho_{0} + AB_{x}^{2})^{-1},
\end{equation}
where $\sigma_{0}$ and $\rho_{0}$ are the zero-field conductivity and resistivity at that temperature, respectively and \textit{C$_{w}$} is a temperature dependent parameter originating from chiral anomaly. The term ($\sigma$$_{0}$ + \textit{a}$\sqrt{B_{x}}$), takes care of the low-field minima in the conductivity, which is generally described as the effect of weak anti-localization in Dirac systems (13,32), while the second term on the right-hand side includes the contribution from the non-linear bands near the Fermi level. Fig. 2\textit{D} illustrates the good agreement between the theoretical expression and experimental data. From Fig. 2\textit{C}, it can be seen that the MR becomes positive at high field which is due to small misalignment of \textit{\textbf{E}} and \textit{\textbf{B}}. The LMR at all temperatures, can be well described using a misalignment angle $\sim$ 2$^{\circ}$. The details are provided in the SI (Fig. S5\textit{A}). We have also measured LMR with \textit{\textbf{E}}$\parallel$\textit{\textbf{B}}, along different crystallographic directions and on several crystals. For all the cases, similar negative MR has been observed, which confirms that negative LMR is associated with \textit{\textbf{E}}$\parallel$\textit{\textbf{B}} configuration rather than any particular crystallographic direction. In Fig. S5\textit{B}, the measured LMR for \textit{\textbf{E}}$\parallel$\textit{\textbf{B}}$\parallel$\textbf{b} axis, is shown as a representative (See SI). Negative LMR has also been observed in few systems other than Dirac or Weyl semimetals. However, in these systems the origin and nature of negative MR are completely different from chiral anomaly (See SI).

\textbf{SdH oscillation and Fermi surface properties.} Another interesting feature which emerges from the transport measurement is the presence of SdH oscillation traceable at field even below 2 T and temperature up to 20 K. This not only gives an insight into the nature of the Fermi surface, but also provides an evidence of very high mobility of the associated charge carriers. From the transverse MR data, it is clear that there are more than one frequency. To extract the oscillatory component $\Delta$$\rho$(\textit{B}), a smooth background is subtracted from $\rho$(\textit{B}). To deconvolute the two components of oscillation, the background subtraction has been done in two steps.
\begin{figure}
\includegraphics[width=0.45\textwidth]{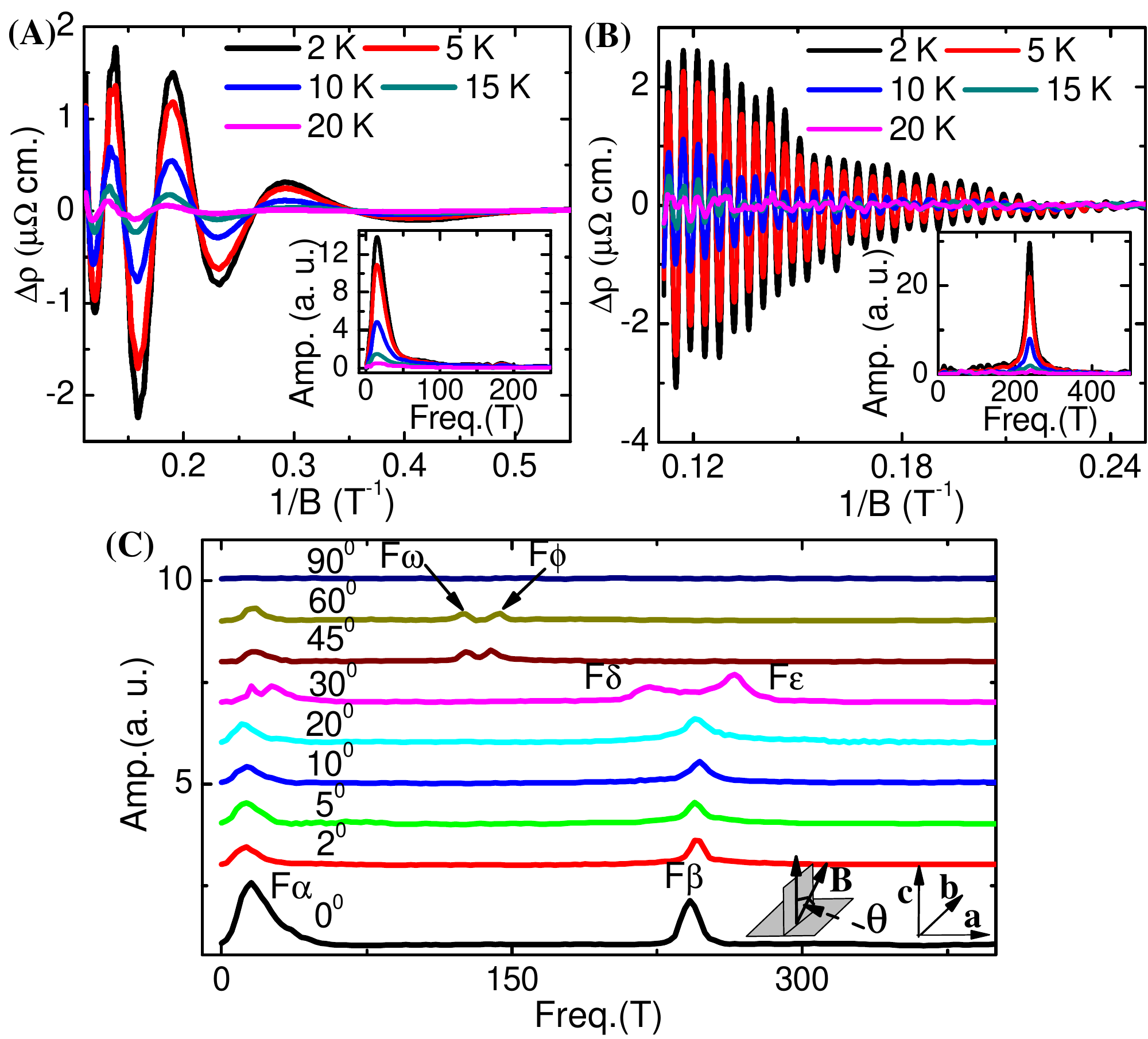}
\caption{\textit{(A)} \& \textit{(B)} SdH oscillation obtained by subtracting smooth background from magnetoresistance measurement, plotted with inverse magnetic field (1/\textit{B}) at different temperatures for the two deconvoluted components. The insets in both figures show the corresponding FFT results. \textit{(C)} The angle dependence of oscillation frequencies. For clarity, FFT results for different angles are shifted vertically. The schematic of the experimental setup is shown in the inset.}
\end{figure}
In Fig. 3\textit{A} and Fig. 3\textit{B}, $\Delta$$\rho$(\textit{B}), for two different components, is plotted as a function of 1/\textit{B} at several representative temperatures. As the oscillation peaks are very sharp and the field interval used in the measurements is not too small compared to the peak width, some fluctuations in the intensity have been observed in Fig. 3\textit{B}. Using a much smaller field interval, we have observed that the peak intensity becomes systematic (See SI Fig. S6\textit{A}). The fast Fourier transform (FFT) analysis of the oscillatory components reveals oscillation frequencies 14 T and 238 T. The obtained frequencies indicate the existence of a very large and a small Fermi surface cross-sections perpendicular to \textbf{c} axis. Using the Onsager relationship, $F = (\varphi_{0}/2\pi^{2}) A_{F}$, where $\varphi$$_{0}$ is the single magnetic flux quantum and A$_{F}$ is the Fermi surface cross-section perpendicular to the applied magnetic field, we have calculated cross-sections 1.4$\times$10$^{- 3}$ ${\AA}$$^{- 2}$ and 22.7$\times$10$^{- 3}$ ${\AA}$$^{- 2}$ for 14 T and 238 T frequencies, respectively. In Fig. 4\textit{A}, the oscillation amplitude for both the frequencies is shown as a function of temperature and has been fitted using the thermal damping factor of Lifshitz-Kosevich formula, $R_{T} = (2\pi^{2}k_{B}T/\beta)/\sinh(2\pi^{2}k_{B}T/\beta)$, where $\beta$ = $e \hbar B/m^{\ast}$. From the fitting parameters, the cyclotron effective masses (\textit{m$^{\ast}$}) of the charge carriers are determined to be $\sim$ 0.14 \textit{m$_{0}$} and $\sim$ 0.1 \textit{m$_{0}$} for 238 T and 14 T frequencies respectively, where \textit{m$_{0}$} is the rest mass of the free electron. To determine the approximate value of the carrier density, we have employed its relation with the oscillation frequency (33), $\Delta \left(\frac{1}{B}\right) = \frac{2e}{\hbar} \left(\frac{g_{s}g_{v}}{6\pi^{2}n_{3D}}\right)^{2/3}$, where \textit{g$_{s}$} and \textit{g$_{v}$} are the spin and valley degeneracies. We found the carrier densities (\textit{n$_{3D}$}) to be 2$\times$10$^{19}$ cm$^{-3}$ and 3$\times$10$^{17}$ cm$^{-3}$ for the large and small Fermi pockets, respectively. From the magnetic field-induced damping of oscillation amplitude, $\Delta\rho$ $\propto$ $exp(-2 \pi^{2} k_{B} m^{\ast} T_{D}/\hbar e B)$, the Dingle temperatures (\textit{T${_D}$}) are determined to be 11.2 K and 3.4 K for the large and small Fermi pockets, respectively, at 2 K. To get a quantitative estimate about the mobility of the charge carriers in the system, we have calculated the quantum mobility, $\mu_{q}$ = $(e \hbar/2 \pi k_{B} m^{\ast} T_{D})$. The obtained values $\sim$ 1.3$\times$10$^{3}$ cm$^{2}$ V$^{-1}$ s$^{-1}$ and $\sim$6.2$\times$ 10$^{3}$ cm$^{2}$ V$^{-1}$ s$^{-1}$ for the large and small frequencies respectively, imply the significant difference between the mobilities of the carriers, which is expected due to different effective masses of the carriers associated with the Fermi pockets (mobility $\propto \frac{1}{m^{\ast}})$. The quantum mobility in a system is always lower than the classical Drude mobility ($\mu_{c}$), as $\mu_{q}$ is sensitive to both large and small angle scattering, while $\mu_{c}$ is sensitive to only large angle scattering (34). The extracted parameters from SdH oscillation are summarized in Table I. With magnetic field along \textbf{b} axis, no clear oscillation has been recorded up to 9 T applied field. This may be due to heavier effective mass and low mobility of the charge carriers along that direction and/or quasi two-dimensional nature of the Fermi surface in ZrSiS.
\begin{figure}
\includegraphics[width=0.45\textwidth]{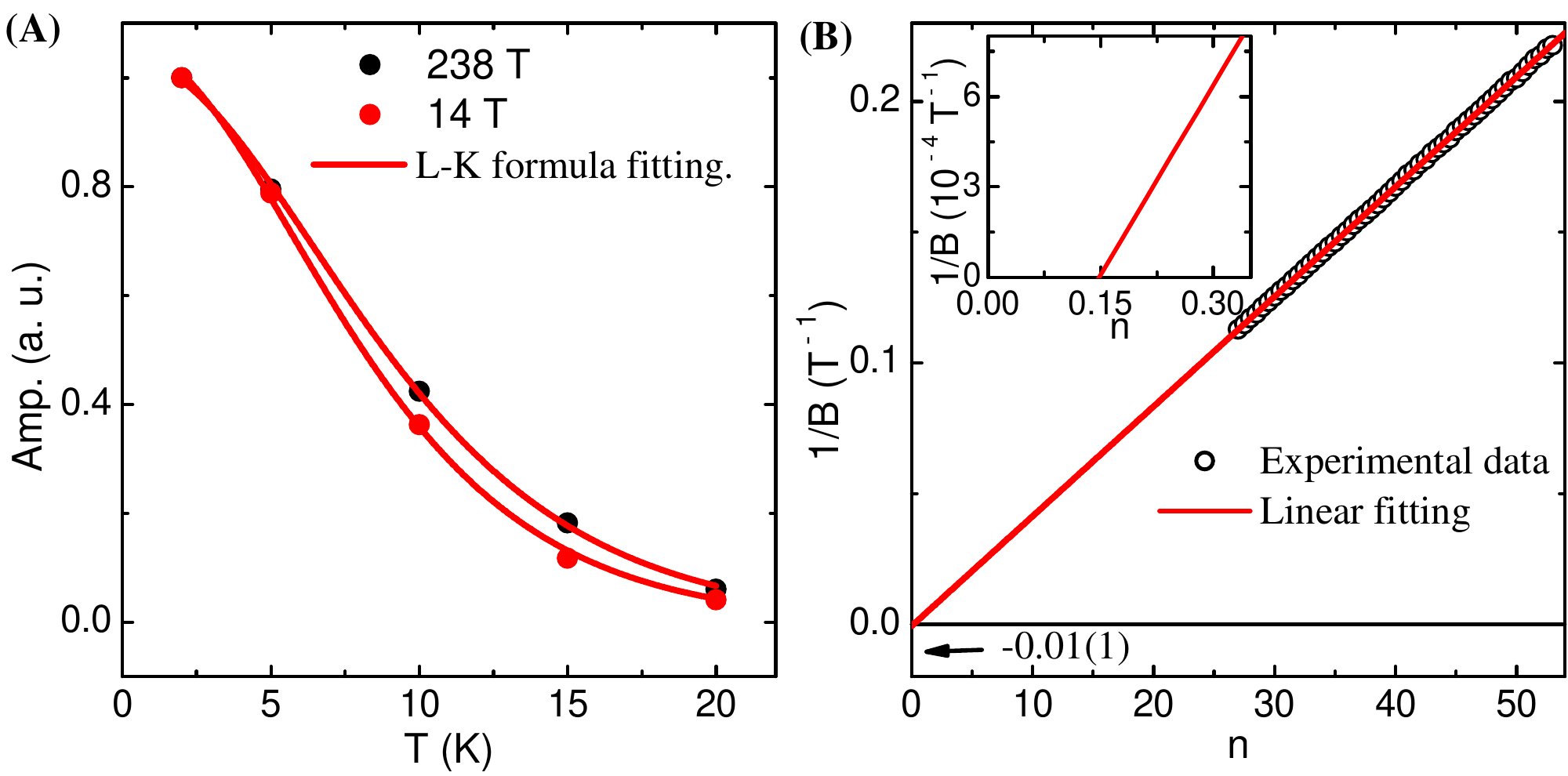}
\caption{ \textit{(A)} Temperature dependence of relative amplitude of SdH oscillation for both the Fermi pockets. \textit{(B)} Landau level index plot for 238 T frequency oscillation. Inset shows the x-axis intercept by extrapolated linear fitting. The value of y-axis intercept is shown by the arrow.}
\end{figure}

\textbf{TABLE I:} Parameters extracted from SdH oscillation for two Fermi pockets (34). \textit{k}$_{F}$, \textit{v}$_{F}$, \textit{l} are the Fermi momentum, Fermi velocity and mean free path of the charge carriers, respectively.
\begin{center}
\begin{tabular}{||c c c c c c c||}
\hline
F & \textit{k}$_{F}$ & m$^{\ast}$ & \textit{v}$_{F}$ & \textit{l} & $\mu_{q}$ & n$_{q}$\\
T & 10 $^{-2}$${\AA}^{-1}$ & m$_{0}$ & 10$^{5}$m/s & nm & 10$^{3}$cm$^{2}$/Vs & 10$^{17}$cm$^{-3}$\\
\hline
238 & 8.5 & 0.14 & 2.4 & 25.7 & 1.3 & 200\\

14 & 2 & 0.1 & 6.9 & 247.4 & 6.2 & 3\\
\hline
\end{tabular}
\end{center}

\textbf{Angle dependence of oscillation frequencies.} For deeper understanding of the Fermi surface geometry, we have performed angle resolved transverse MR measurements and SdH oscillation analysis. The resultant FFT spectra for different directions, are shown in Fig. 3\textit{C}. The inset illustrates the experimental set-up with current along \textbf{a} axis and magnetic field rotated in the \textbf{bc} plane. As illustrated in Fig. 3\textit{C}, the low-frequency component F$_{\alpha}$ (14 T) remains invariant with increasing angle up to 20$^{\circ}$, after which it bifurcates into two very closely spaced frequency components. However, with further increase in angle, they come close to each other and merge to become a single frequency. On the other hand, the high-frequency component F$_{\beta}$ (238 T) splits into two components (F$_{\delta}$ and F$_{\varepsilon}$), which are well separated in the frequency spectra. While the higher one (F$_{\varepsilon}$) among the two, disappears above a certain angle, F$_{\delta}$ is seen to shift towards lower value and then bifurcates (F$_{\omega}$ and F$_{\phi}$). As already discussed, with current along \textbf{a} and magnetic field along \textbf{b} axis ($\theta$ = 90$^{\circ}$), no clear oscillatory component has been found.

\textbf{Berry phase and Zeeman splitting.} In an external magnetic field, a closed orbit is quantized following the Lifshitz-Onsager quantization rule (35), $A_{F}\frac{\hbar}{eB} = 2\pi(n + \frac{1}{2} - \beta - \delta) = 2\pi(n + \gamma - \delta)$, where 2$\pi$$\beta$ is the Berry's phase and $\delta$ is a phase shift determined by the dimensionality, having value 0 and $\pm$ 1/8 for 2D and 3D cases, respectively. The nature of the electronic band dispersion is determined by the value of the Berry phase, which is 0 for the conventional metals with parabolic band dispersion and $\pi$ for the Dirac/Weyl type electronic system with linear band dispersion. The quantity $\gamma - \delta = \frac{1}{2} - \beta - \delta$, can be extracted from the x-axis (along which the Landau level index \textit{n} has been plotted) intercept in the Landau level fan diagram and takes a value in the range - 1/8 to + 1/8 for 3D Dirac fermions (35). In Fig. 4\textit{B}, the Landau level fan diagram for the larger Fermi pocket in ZrSiS, has been plotted, assigning maxima of the SdH oscillation as integers (\textit{n}) and minima as half-integers (\textit{n} + 1/2). Extrapolated linear fitting gives an intercept of 0.15(3) (shown in the inset). The sharp, symmetric and well-separated oscillation peaks over a wide range (\textit{n} = 27 to 53) and traceable down to $\sim$ 4 T, imply no significant error in determining the value of the intercept from the linear \textit{n} vs. 1/\textit{B} fit. On the other hand, with higher magnetic field to achieve lower Landau level, the non-linearity in the index plot may arise due to the Zeeman splitting of oscillation peaks as observed for 14 T frequency and discussed below. Similar to that observed for 238 T frequency, a small intercept $\sim$ -0.01 is obtained for 14 T frequency and shown in Fig. S6\textit{B} (See SI). For both the Fermi pockets, the intercepts are very close to the range $\pm$ 1/8. For the smaller frequency, the experimental peak positions are seen to deviate slightly from a straight line which can be attributed to the Zeeman splitting of the Landau levels (34,36). Although the presence of Zeeman splitting is not clearly visible in the SdH oscillation, the spin-split peaks can be easily distinguished in the de Haas-van Alphen (dHvA) oscillation in our magnetization measurements (Fig. S7\textit{A} and S7\textit{B}) (See SI). Taking the peak and valley positions of the lower field oscillations, which are almost free from the Zeeman splitting, we have also plotted the Landau level fan diagram for smaller frequency from dHvA oscillation (Fig. S7\textit{C}) and obtained a small intercept 0.05(2). Furthermore, we have calculated the Berry's phase from the SdH oscillations at different angles (up to 20$^{\circ}$) and did not find any significant change. Finding reasonably accurate value of Berry's phase for higher angles is much more complicated due to the presence of multiple oscillation frequencies.

\textbf{Hall measurement.} To determine the nature of the charge carriers of two Fermi pockets, the Hall effect measurement has been performed. At 300 K, the Hall resistivity is found to be almost linear with field and positive (Fig. 5\textit{A}), which indicate holes as majority carriers, consistent with the earlier ARPES report (20). With decreasing temperature, the Hall resistivity develops a sublinear character and at around 50 K, it changes sign from positive to negative at high magnetic field, confirming the existence of more than one type of carrier. The overall behavior of the Hall resistivity can be explained by considering low-mobility holes and higher-mobility electrons associated with large and small Fermi pockets, respectively. Following the classical two-band model (37), the Hall resistivity is fitted in Fig. S8 (See SI). Obtained electron and hole densities, 1.6$\times$10$^{17}$ cm$^{-3}$ and 6$\times$10$^{19}$ cm$^{-3}$, respectively, are in agreement with those calculated from SdH oscillation. As expected, at 5 K, large electron mobility $\sim$ 2$\times$10$^{4}$ cm$^{2}$ V$^{-1}$ s$^{-1}$ and hole mobility $\sim$ 2.8$\times$10$^{3}$ cm$^{2}$ V$^{-1}$ s$^{-1}$ have been obtained from the fitted parameters. From the Hall resistivity, it is clear that at least two band crossings are present in the electronic band structure of ZrSiS at different energy values, as shown schematically in Fig. 5\textit{B}.
\begin{figure}
\includegraphics[width=0.5\textwidth]{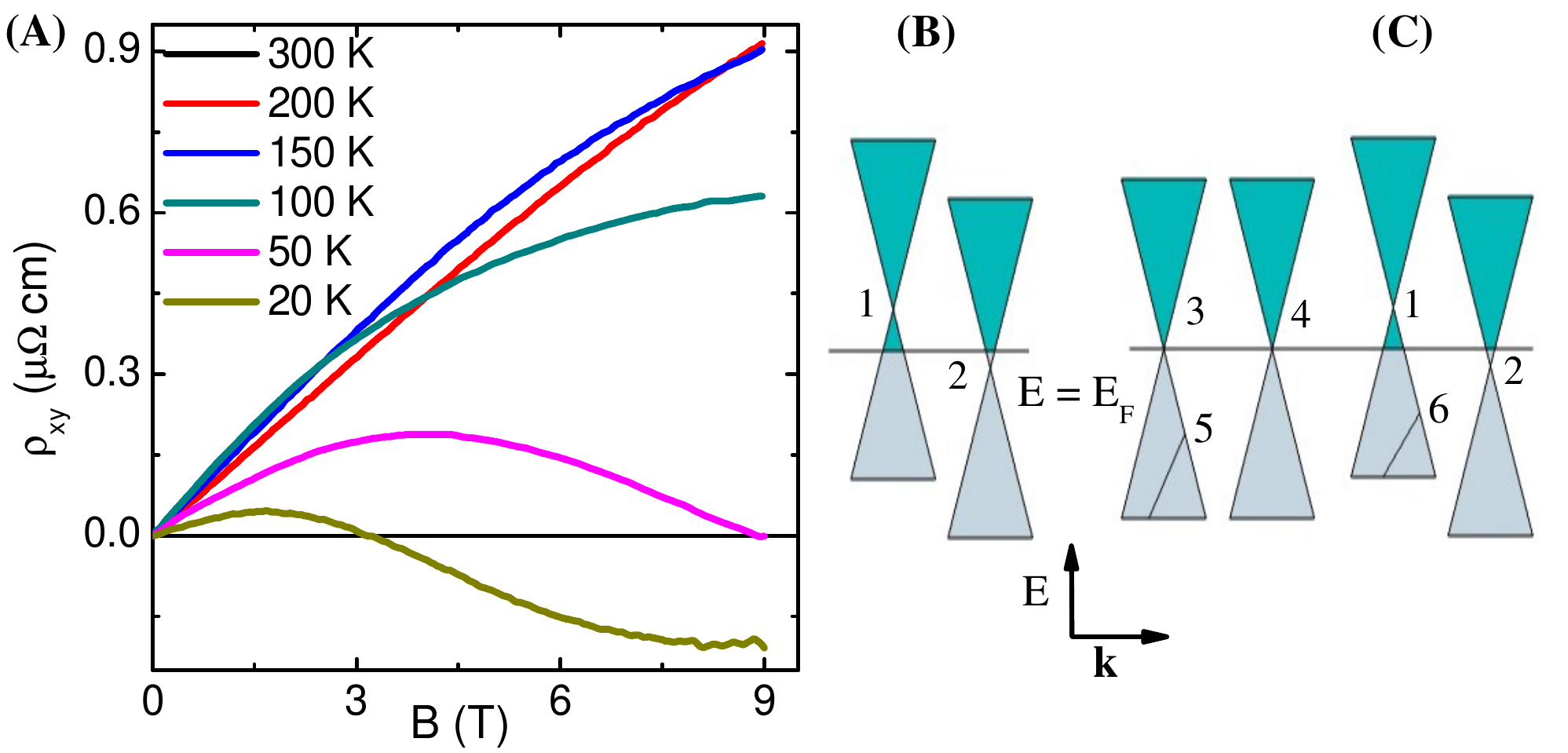}
\caption{\textit{(A)} Field dependence of the Hall resistivity measured at different temperatures. \textit{(B)} Schematic explaining transport measurement results. \textit{(C)} Schematic illustrating multiple Dirac cones in ZrSiS as described in earlier reports (19,20).}
\end{figure}
The earlier reports on ARPES and band structure calculations suggest the presence of multiple Dirac crossings at different energy values as illustrated in the schematic Fig. 5\textit{C}. As shown, the Dirac cones 1 and 2 cross the Fermi energy, having Dirac points at two different energy values. Among the rest, 3 and 4 have their band crossing points almost at the chemical potential with negligible Fermi surface, while 5 and 6 are lying well below the Fermi energy. Thus in this configuration, it is expected that only Dirac cones 1 and 2 will contribute to the transport properties of ZrSiS, which is consistent with our magneto-transport results.

\textbf{Summary}

In conclusion, we present the systematic study of the magneto-electronic transport properties on ZrSiS single crystals. Magnetic field-induced metal-semiconductor-like crossover along with strongly anisotropic transport properties have been observed. The anisotropic MR along different crystallographic axes is in good agreement with the quasi two-dimensional nature of the Fermi surface observed in ARPES. Transverse MR approaches an extremely large value $\sim$ 1.4 $\times$ 10$^{5}$ \% at 2 K and 9 T, without any sign of saturation. Under parallel \textit{\textbf{E}} and \textit{\textbf{B}} configuration, the observed negative MR implies Adler-Bell-Jackiw chiral anomaly of three-dimensional Dirac fermions in ZrSiS. The SdH oscillation reveals two inequivalent Fermi surface cross-sections perpendicular to crystallographic \textbf{c} axis. The Dirac fermionic nature of the charge carriers is also confirmed from the observed non-trivial $\pi$ Berry phase in Landau level fan diagram for both the Fermi pockets. Non-linear field dependence of Hall resistivity indicates the presence of both electron and hole type charge carriers. Classical two-band fitting of Hall resistivity reveals high mobilities for both types of charge carrier. SdH oscillation along with Hall measurement reflect multiple band crossings at different energy values of the electronic band structure. We believe, the present work not only makes a substantive experimental contribution in this contemporary area of research but also can encourage further extensive works in ZrSiS and other members of the family.

\textit{Note added:} During the submission process of our manuscript, several reports on magnetotransport and magnetization measurements in ZrSiS were submitted to arXiv.org (38-40), supporting the major conclusions of our work.

\textbf{Materials and Methods}

{\small The single crystals were grown by standard iodine vapor transport technique and characterized using HRTEM and EDX. The transport measurements were performed in PPMS (Quantum Design) and cryogen free system (Cryogenic) via four-probe technique. Magnetic measurements were done in MPMS3 (Quantum Design). See \textbf{SI Materials and Methods} for details.}

\textbf{Acknowledgements:} {\small We thank N. Khan, A. Paul and S. Roy for their help during measurements and useful discussions.}\\

{\scriptsize 1. Hasan MZ, et al. (2010) \textit{Colloquium:} Topological insulators. \textit{Rev Mod Phys} 82(4): 3045-3067.

2. Liu ZK, et al. (2014) A stable three-dimensional topological Dirac semimetal Cd$_{3}$As$_{2}$. \textit{Nat Mat} 13(7): 677-681.

3. Xu SY, et al. (2015) Discovery of a Weyl fermion semimetal and topological Fermi arcs. \textit{Science} 349(6248): 613-617.

4. Daughton JM (1999) GMR applications. \textit{J Magn Magn Mater} 192(2): 334-342.

5. Wolf SA, et al. (2001) Spintronics: A spin-based electronics vision for the future. \textit{Science} 294(5546): 1488-1495.

6. Lenz J (1990) A review of magnetic sensors. \textit{Proc IEEE} 78(6): 973-989.

7. Wang Z, et al. (2012) Dirac semimetal and topological phase transitions in A$_{3}$Bi (A=Na, K, Rb). \textit{Phys Rev B} 85(19): 195320.

8. Wang Z, et al. (2013) Three-dimensional Dirac semimetal and quantum transport in Cd$_{3}$As$_{2}$. \textit{Phys Rev B} 88(12): 125427.

9. Neupane M, et al. (2014) Observation of a three-dimensional topological Dirac semimetal phase in high-mobility Cd$_{3}$As${2}$. \textit{Nat Commun} 5: 3786.

10. Liang T, et al. (2015) Ultrahigh mobility and giant magnetoresistance in the Dirac semimetal Cd$_{3}$As${2}$. \textit{Nat Mat} 14(3): 280-284.

11. Liu ZK, et al. (2015) Discovery of a Three-Dimensional Topological Dirac Semimetal, Na$_{3}$Bi. \textit{Science} 343(6173): 864-867.

12. Weng H, et al. (2015) Weyl Semimetal Phase in Noncentrosymmetric Transition-Metal Monophosphides. \textit{Phys Rev X} 5(1): 011029.

13. Huang X, et al. (2015) Observation of the Chiral-Anomaly-Induced Negative Magnetoresistance in 3D Weyl Semimetal TaAs. \textit{Phys Rev X} 5(3): 031023.

14. Shekhar C, et al. (2015) Extremely large magnetoresistance and ultrahigh mobility in the topological Weyl semimetal candidate NbP. \textit{Nat Phys} 11(8): 645-649.

15. Borisenko S, et al. (2015) Time-Reversal Symmetry Breaking Type-II Weyl State in YbMnBi$_{2}$. arXiv:1507.04847.

16. Kim Y, et al. (2015) Dirac Line Nodes in Inversion-Symmetric Crystals. \textit{Phys Rev Lett} 115(3): 036806.

17. Bian G, et al. (2016) Drumhead surface states and topological nodal-line fermions in TlTaSe$_{2}$. \textit{Phys Rev B} 93(12): 121113(R).

18. Bian G, et al. (2016) Topological nodal-line fermions in spin-orbit metal PbTaSe$_{2}$. \textit{Nat Commun} 7: 10556.

19. Xu Q, et al. (2015) Two-dimensional oxide topological insulator with iron-pnictide superconductor LiFeAs structure. \textit{Phys Rev B} 92(20): 205310.

20. Schoop LM, et al. (2016) Dirac cone protected by non-symmorphic symmetry and three-dimensional Dirac line node in ZrSiS. \textit{Nat Commun} 7: 11696.

21. Zhao Y, et al. (2015) Anisotropic Fermi Surface and Quantum Limit Transport in High Mobility Three-Dimensional Dirac Semimetal Cd$_{3}$As$_{2}$. \textit{Phys Rev X} 5(3): 031037.

22. Ziman JM (2001) \textit{Electrons and Phonons}. Classics Series, Oxford University Press, New York.

23. Tafti FF, et al. (2016) Resistivity plateau and extreme magnetoresistance in LaSb. \textit{Nat Phys} 12(3): 272-277.

24. Sun S, et al. (2016) Large magnetoresistance in LaBi: origin of field-induced resistivity upturn and plateau in compensated semimetals. arXiv:1601.04618v1.

25. Ali MN, et al. (2014) Large, non-saturating magnetoresistance in WTe$_{2}$. \textit{Nature} 514(7521): 205.

26. Wang K, et al. (2014) Anisotropic giant magnetoresistance in NbSb$_{2}$. \textit{Sci Rep} 4: 7328.

27. Kopelevich Y, et al. (2006) Universal magnetic-field-driven metal-insulator-metal transformations in graphite and bismuth. \textit{Phys Rev B} 73(16): 165128.

28. Wang YL, et al. (2015) Origin of the turn-on temperature behavior in WTe$_{2}$. \textit{Phys Rev B} 92(18): 180402(R).

29. Hu J, et al. (2008) Classical and quantum routes to linear magnetoresistance. \textit{Nat Mat} 7(9): 697-700.

30. McKenzie RH, et al. (1998) Violation of Kohler's rule by the magnetoresistance of a quasi-two-dimensional organic metal. \textit{Phys Rev B} 57(19): 11854.

31. Li H, et al. (2016) Negative magnetoresistance in Dirac semimetal Cd$_{3}$As$_{2}$. \textit{Nat Commun} 7: 10301.

32. Xiong J, et al. (2015) Evidence for the chiral anomaly in the Dirac semimetal Na$_{3}$Bi. \textit{Science} 350(6259): 413-416.

33. Shoenberg D, (1984) \textit{Magnetic oscillations in metals}. Cambridge Univ. Press.

34. Narayanan A, et al. (2015) Linear Magnetoresistance Caused by Mobility Fluctuations in n-Doped Cd$_{3}$As$_{2}$. \textit{Phys Rev Lett} 114(11): 117201.

35. Murakawa H, et al. (2013) Detection of Berry's Phase in a Bulk Rashba Semiconductor. \textit{Science} 342(6165): 1490-1493.

36. Taskin AA, et al. (2011) Berry phase of nonideal Dirac fermions in topological insulators. \textit{Phys Rev B} 84(3): 035301.

37. Hurd CM (1972) \textit{The Hall effect in metals and alloys}. Plenum Press, New York.

38. Ali MN, et al. (2016) Butterfly Magnetoresistance, Quasi-2D Dirac Fermi Surfaces, and a Topological Phase Transition in ZrSiS. arXiv:1603.09318v2.

39. Wang X, et al. (2016) Evidence of both surface and bulk Dirac bands in ZrSiS and the unconventional magnetoresistance. arXiv:1604.00108.

40. Hu J, et al. (2016) Evidence of Dirac cones with 3D character probed by dHvA oscillations in nodal-line semimetal ZrSiS. arXiv:1604.01567.

41. Haneveld AK, et al. (1964) Zirconium silicide and germanide chalcogenides preparation and crystal structures. \textit{Recueil des Travaux Chimiques des Pays-Bas} 83(8): 776-783.

42. Tafti FF, et al. (2016) Temperature-field phase diagram of extreme magnetoresistance. \textit{Proc Natl Acad Sci USA} 113(25): E3475.

43. Kikugawa N, et al. (2016) Interplanar coupling-dependent magnetoresistivity in high-purity layered metals. \textit{Nat Commun}, 7: 10903.

44. Ando Y (2013) Topological Insulator Materials. \textit{J Phys Soc Jpn} 82(10): 102001.

45. Huynh KK, et al. (2011) Both Electron and Hole Dirac Cone States in Ba(FeAs)$_{2}$ Confirmed by Magnetoresistance. \textit{Phys Rev Lett} 106(21): 217004.}

\newpage

\begin{center}
\textbf{Supplementary information: Large nonsaturating magnetoresistance and signature of non-degenerate Dirac nodes in ZrSiS}
\end{center}

\setcounter{figure}{0}

\textbf{SI Materials and Methods.} Single crystals of ZrSiS were grown in two steps via iodine vapor transport. At first, the polycrystalline powder was synthesized using elemental Zr (Alfa Aesar 99.9\%), Si (Strem Chem. 99.999\%) and S (Alfa Aesar 99.9995\%). The details are described elsewhere (41). Then the polycrystalline powder together with iodine in a concentration of 5 mg/cm$^{3}$ were sealed in a 20 cm long quartz tube under vacuum. The quartz tube was kept in a gradient furnace for 72 h with the powder at 1100$^0$C, while the cooler end at 1000$^0$C. Shinny rectangular plate-like crystals were obtained at the cooler end. High-resolution transmission electron microscopy (HRTEM) of the grown crystals has been done in FEI, TECNAI G$^{2}$ F30, S-TWIN microscope operating at 300 kV equipped with a GATAN Orius SC1000B CCD camera. Energy-dispersive X-ray (EDX) spectroscopy of the grown ZrSiS crystals has been performed using the same microscope with scanning unit and high-angle annular dark-field scanning (HAADF) detector from Fischione (Model 3000). Transport measurements were performed via four-probe technique in a 9 T Physical Property Measurement System (Quantum Design) in ac transport option as well as in a 9 T cryogen free measurement system (Cryogenic) using nanovoltmeter (Keithley). Magnetic measurements were done in a 7 T SQUID-VSM MPMS 3 (Quantum Design).\\

\textbf{SI Sample characterization.} In Fig. S1\textit{A}, a single crystal of ZrSiS with typical dimensions 1.8 mm $\times$ 1 mm $\times$ 0.14 mm, is shown with different crystallographic axes. The crystals cleave perpendicular to \textbf{c} axis similar to earlier report (20). High-resolution transmission electron microscopy (HRTEM) image of the crystal along \textbf{ac} plane (Fig. S1\textit{B}) confirms the high quality crystalline nature and the layered structure of the lattice with $\sim$8 ${\AA}$ interlayer distance. The electron diffraction patterns obtained in HRTEM are shown in Fig. S1\textit{C} and S1\textit{D} with corresponding Miller indices of the lattice planes. Energy-dispersive X-ray spectroscopy (Fig. S2) verifies almost perfect stoichiometry and absence of any impurity in the grown crystals. Magnetotransport measurements were done on several crystals from the same batch, which reproduced similar results.\\

\begin{figure}
\includegraphics[width=0.5\textwidth]{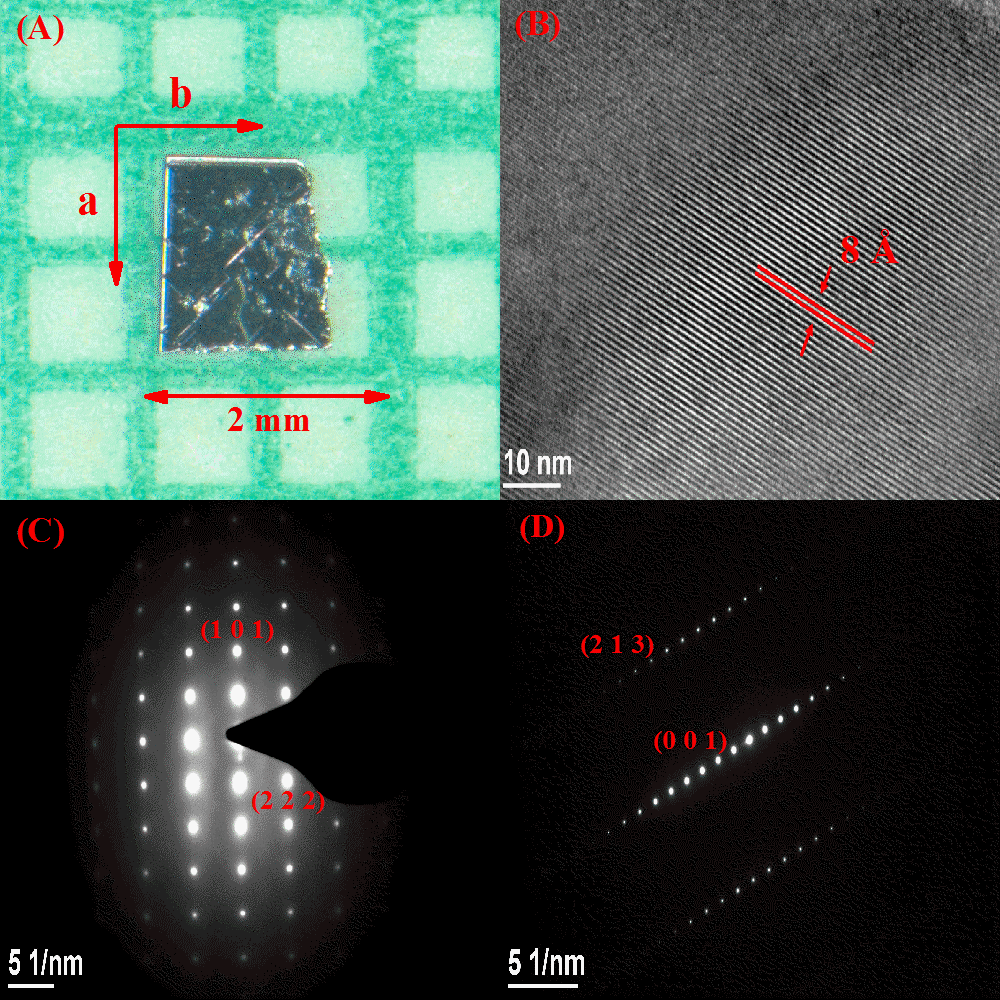}
\renewcommand{\figurename}{Fig.S}
\caption{\textit{(A)} ZrSiS single crystal with different crystallographic directions. \textit{(B)} HRTEM image along \textbf{ac}-plane. \textit{(C)} and \textit{(D)} Selected area electron diffraction (SAED) pattern obtained through HRTEM measurement.}
\end{figure}

\begin{figure}
\includegraphics[width=0.5\textwidth]{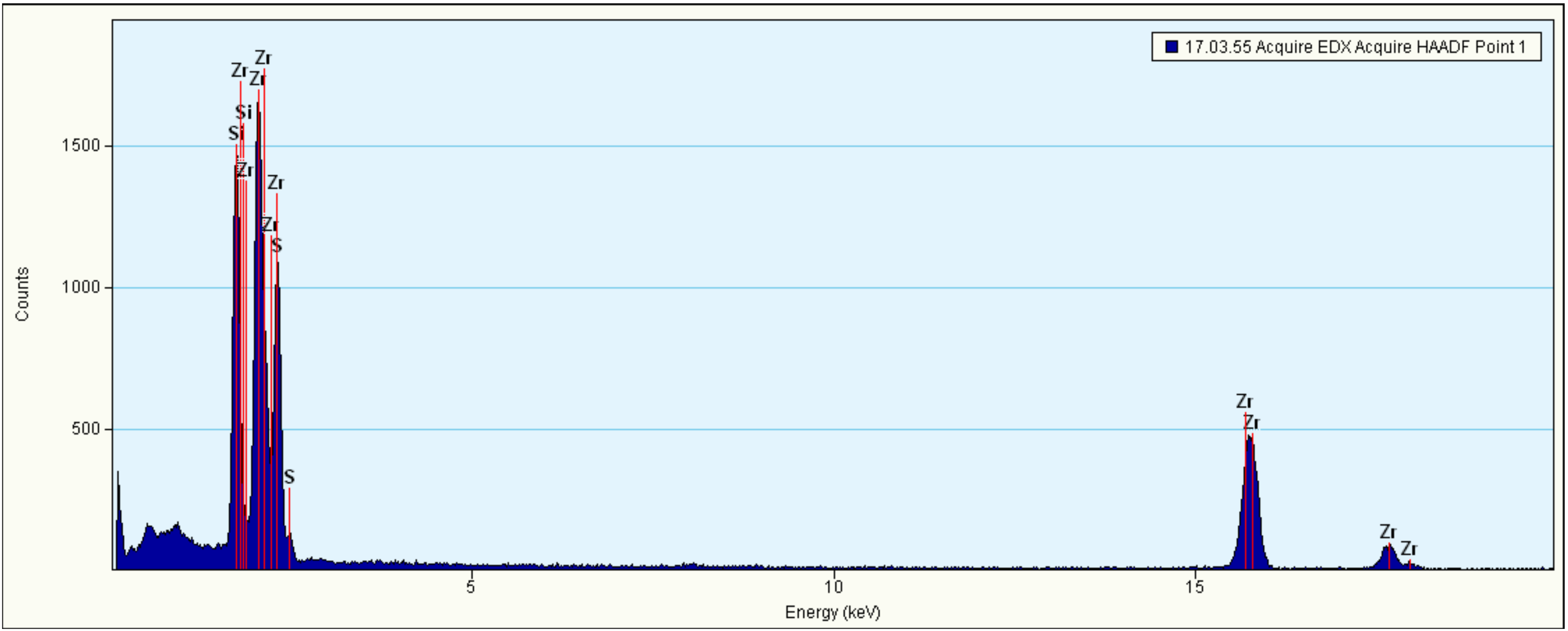}
\renewcommand{\figurename}{Fig.S}
\caption{Single crystal EDX spectroscopy data.}
\end{figure}

\textbf{Magnetotransport properties.} The zero-field resistivity shows two regions with different temperature dependence. As shown in the Fig. S3\textit{A}, the resistivity obeys \textit{T}$^{3}$ dependence at low temperature, which is followed by almost linear dependence at high temperature.

\begin{figure}
\includegraphics[width=0.5\textwidth]{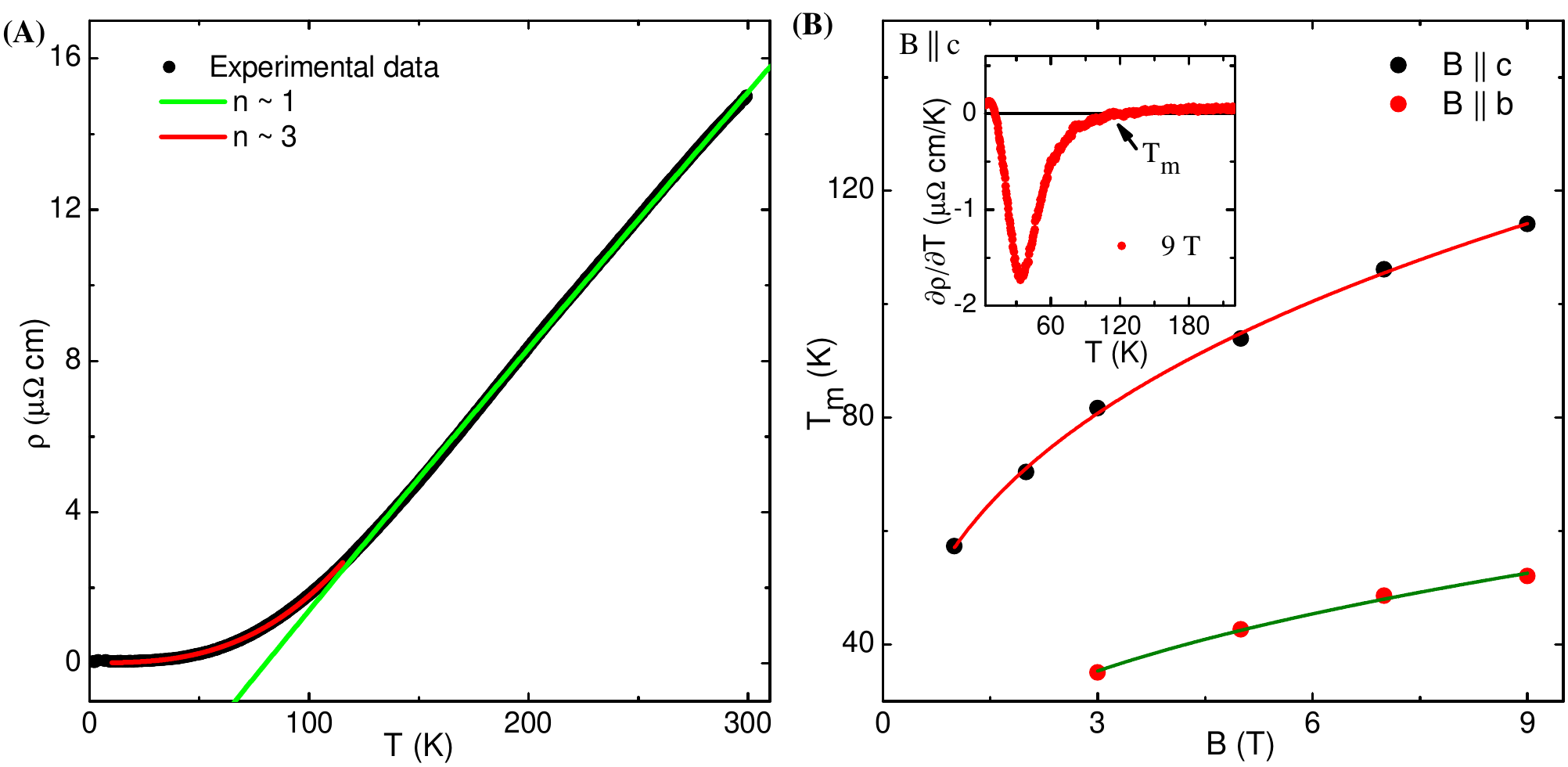}
\renewcommand{\figurename}{Fig.S}
\caption{\textit{(A)} Temperature dependence of zero-field resistivity. The experimental data is fitted using $\rho(T) = \rho_{0} + AT^{n}.$ \textit{(B)} Field dependence of metal-semiconductor crossover temperature for two field directions. Inset shows the temperature dependence of $\partial$$\rho$/$\partial$T at a field of 9 T, applied parallel to \textbf{c} axis. The point, where the curve crosses the x-axis, indicates the crossover temperature (\textit{T$_{m}$}).}
\end{figure}

Field induced metal-semiconductor crossover is indicated by the increase in resistivity with decreasing temperature in presence of magnetic field. The crossover temperature (\textit{T$_{m}$}) is identified as the temperature where resistivity shows minimum, i.e., the temperature where $\partial$$\rho$/$\partial$T becomes zero. The magnetic field dependence of \textit{T$_{m}$} is shown in Fig. S3\textit{B}. For both the applied field directions, \textit{T$_{m}$} is seen to be $\propto (B - B_{0})^{1/3}$.\\

The logarithmic behavior of resistivity with inverse temperature is shown in Figs. S4\textit{A} and \textit{B} for magnetic field applied along two different crystallographic directions. At the point of metal-semiconductor like crossover, the sign of the slope changes. From the slope of the curves at the linear regions, the values of thermal activation energy (\textit{E$_{g}$}) have been calculated. As the curves are linear over a very small temperature range, the calculated gap depends on the region of linear fitting and hence, becomes a function of \textit{T}. Following Tafti \textit{et al.} (42), we have calculated energy gap at different temperature regions. The error bar in Fig. S4\textit{C} represents the maximum change in \textit{E$_{g}$}, when we change the linear fitting range. The calculated energy gap exhibits magnetic field dependence for both the directions of the applied field.
\begin{figure}
\includegraphics[width=0.5\textwidth]{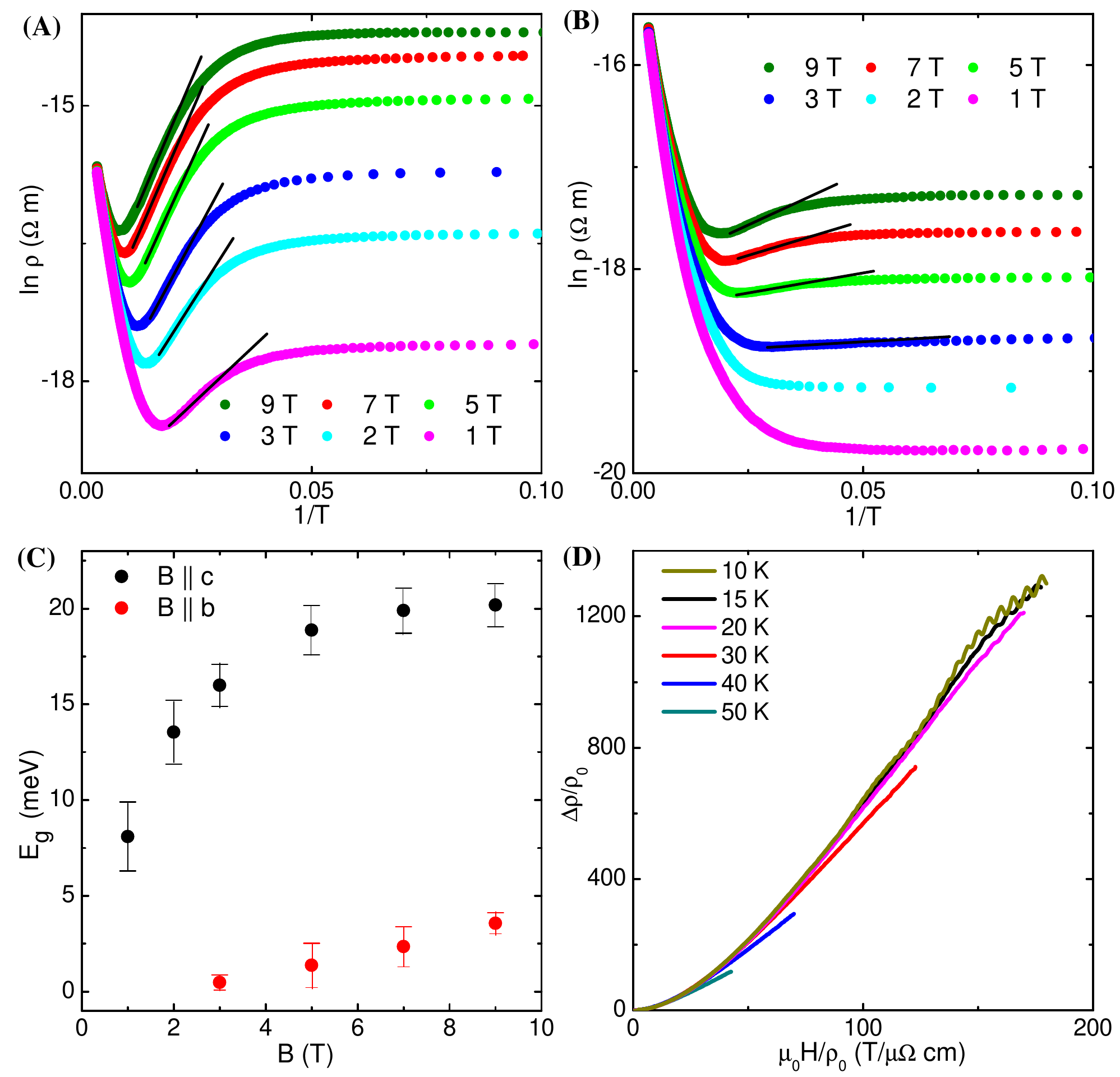}
\renewcommand{\figurename}{Fig.S}
\caption{ln($\rho$) is plotted against T$^{-1}$ at different transverse magnetic fields for \textit{(A)} \textit{\textbf{B}} $\parallel$ \textbf{c} - axis, \textit{(B)} \textit{\textbf{B}} $\parallel$ \textbf{b} - axis. \textit{(C)} Magnetic field dependence of energy gap for two different field directions. Error bars represent the maximum change in \textit{E$_{g}$}, when we change the linear fitting range.\textit{(D)} The Kohler's rule scaling of MR data with field parallel to \textbf{c} axis.}
\end{figure}

As shown in Fig. S4\textit{D}, the magnetoresistance (MR) data have been plotted as a function of $\mu_{0}H/\rho_{0}$. The curves at different temperatures do not merge, indicating the violation of Kohler's rule in ZrSiS.

\begin{figure}
\includegraphics[width=0.5\textwidth]{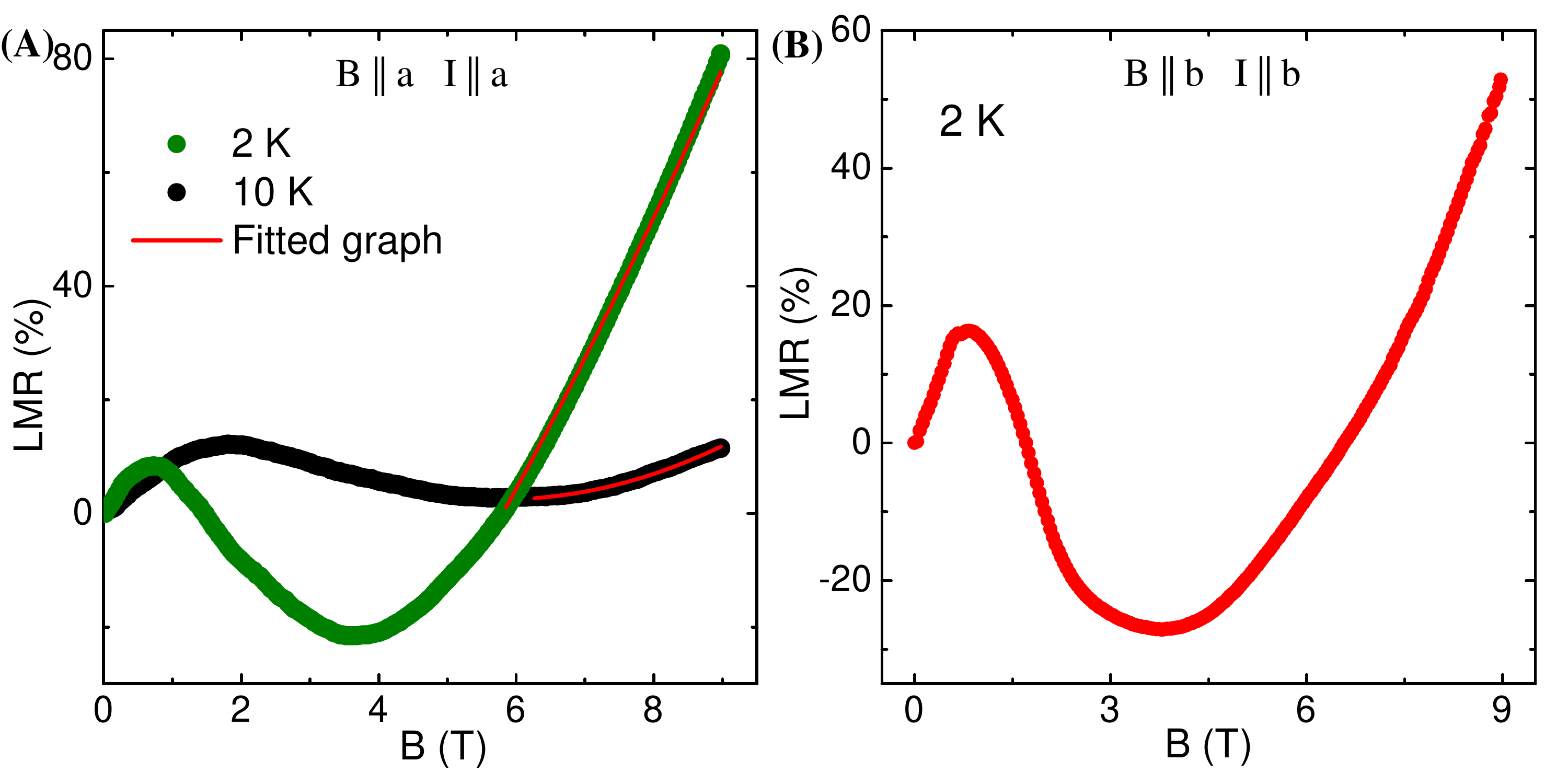}
\renewcommand{\figurename}{Fig.S}
\caption{\textit{(A)} Fitting of high-field region of LMR (\textit{\textbf{B}} $\|$ I $\|$ \textbf{a}) assuming a small misalignment angle between \textit{\textbf{B}} and \textit{\textbf{E}}. \textit{(B)} Longitudinal magnetoresistance (LMR) at 2 K with current and field along \textbf{b} axis.}
\end{figure}

Fig. S5\textit{A} illustrates the LMR of ZrSiS. MR becomes positive at higher fields, which is due to the small misalignment of \textit{\textbf{E}} and \textit{\textbf{B}}. Therefore, a competition between the negative LMR and positive TMR components occurs. As both of these components have different temperature dependence and TMR decreases more rapidly with increasing temperature, the minimum in MR shifts towards higher field with increasing temperature. By fitting the high-field region of MR, a small misalignment angle $\sim$ 2$^{\circ}$ is determined for all the temperatures. Similar behavior has been observed, when \textit{\textbf{E}} and \textit{\textbf{B}} both are applied along arbitrary crystallographic directions in the \textbf{ab} plane. As a representative, in Fig. S5\textit{B} the LMR at 2 K is shown with \textit{\textbf{E}} and \textit{\textbf{B}} along \textbf{b} axis. Besides the Dirac and Weyl semimetals, a few ultraclean layered materials such as PdCoO$_{2}$, PtCoO$_{2}$ and Sr$_{2}$RuO$_{4}$ also show negative LMR (43). However, the nature of the observed negative MR in these compounds is completely different from that originated from chiral anomaly. In these layered materials, negative MR appears, when \textit{\textbf{E}} and \textit{\textbf{B}} are along certain crystallographic direction. For other directions, MR is positive, even for parallel \textit{\textbf{E}} and \textit{\textbf{B}} configuration, unlike Dirac and Weyl semimetals. For example, in PdCoO$_{2}$ and PtCoO$_{2}$, the negative MR is observed only when \textbf{E}$\parallel$\textbf{B}$\parallel$\textbf{c} axis or \textit{\textbf{B}} is close to an Yamaji angle and in Sr$_{2}$RuO$_{4}$ only when \textit{\textbf{E}} and \textit{\textbf{B}} are within 10$^{\circ}$ of the \textbf{c} axis. Moreover, in these systems, the MR decreases linearly with increasing field from its zero-field value. On the other hand, the chiral anomaly induced negative MR in 3D Dirac and Weyl semimetals is quadratic in field as described by Eq. 1 in the main part of the article.\\

As shown in Fig. 3\textit{B}, some fluctuations in the amplitude of the SdH oscillation peaks have been observed for 238 T frequency component. These fluctuations are prominent at 2 K but not so clearly visible for other temperatures. As the oscillation peaks are very sharp and the field interval used in the measurements is not too small compared to the peak width, it is expected that some fluctuations would appear in the peak intensity. The peak intensity becomes systematic, when we measured with much smaller field interval (Fig. S6\textit{A}). However, we have not observed any visible change in peak positions. Moreover, the positions of the fluctuations are completely random and change, when the field interval is changed.\\

The Landau level index plot for the smaller Fermi pocket has been shown in Fig. S6\textit{B}. Due to the weak non-linearity in the index plot, the intercept has been obtained from the best linear fit. Although this method introduces some error in the observed intercept, even with maximum error ($\pm$0.18), which has been calculated from the linear fitting of the first three and last three points, the intercept is close to the previously mentioned theoretical range and far from that expected for the conventional quadratic band (0.5). The small deviation in the experimental data from the straight line is a consequence of the unequally spaced maxima/minima in SdH oscillation (Fig. 3\textit{A}) and likely to occur due to the Zeeman splitting. The spin-split peaks can be clearly seen in the dHvA oscillation (Fig. S7\textit{A}). In Fig. S7\textit{B}, the oscillatory part of magnetic susceptibility $\Delta$$\chi$ = dM/dB, is plotted for the smaller frequency, after subtracting the background. For the Landau level fan diagram, if we assume integer indices \textit{n} for maxima in $\Delta$\textit{M}, the maxima in $\Delta$$\chi$ correspond to \textit{n} + 1/4 (44). Therefore, the Landau level index \textit{n} + 1/4 is plotted in Fig. S7\textit{C} for smaller frequency, taking the peak and valley positions of the lower field oscillations, which are almost free from the Zeeman splitting. The linear nature of the index plot indicates almost no error in the obtained intercept. From Fig. S7\textit{C}, a small intercept 0.05(2) is obtained, which confirms the Dirac fermionic nature of the charge carriers associated to small Fermi pocket.\\

\begin{figure}
\includegraphics[width=0.5\textwidth]{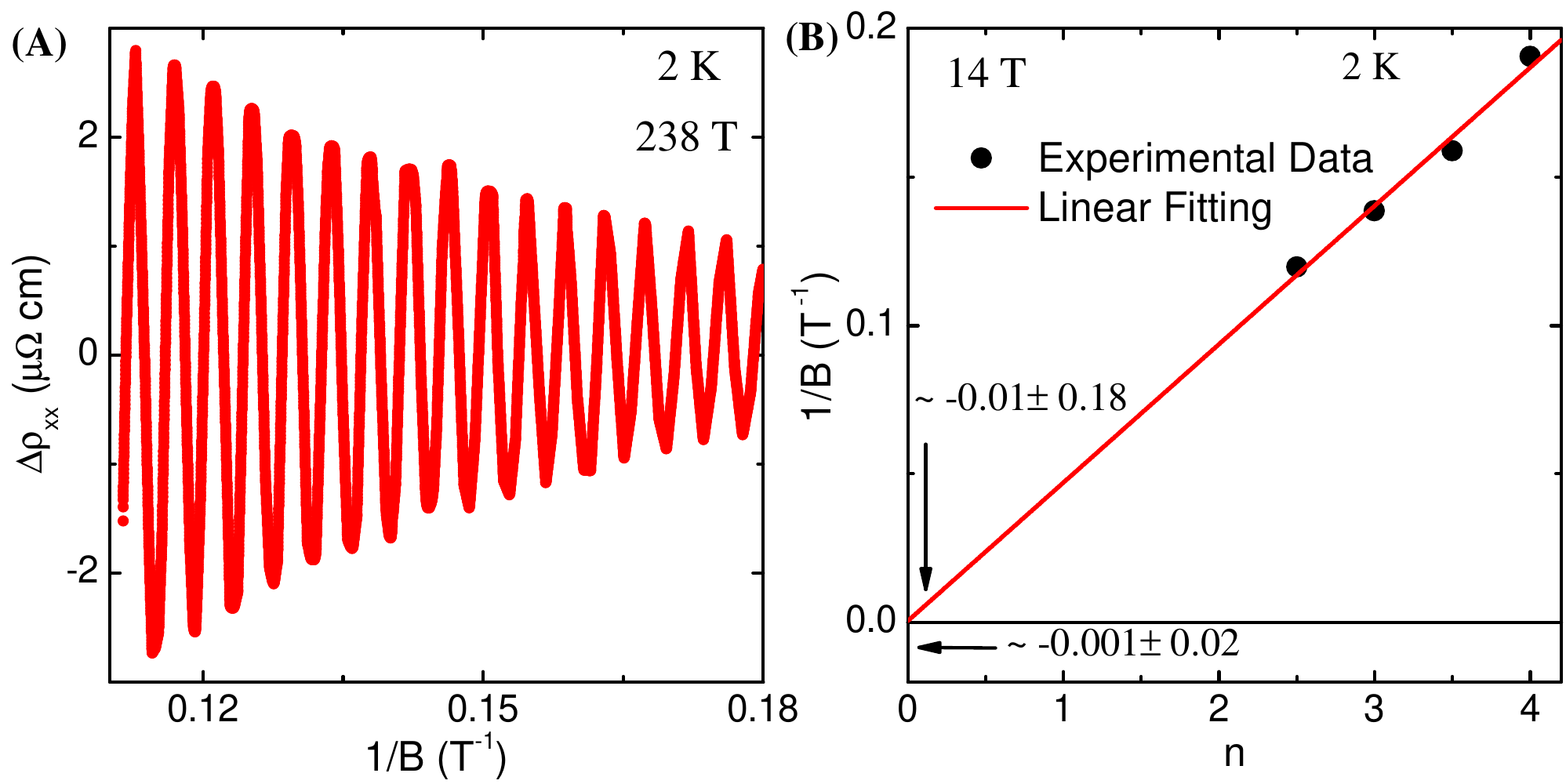}
\renewcommand{\figurename}{Fig.S}
\caption{\textit{(A)} The SdH oscillation at 2 K for 238 T frequency, measured with very small field interval. \textit{(B)} Landau level index plot for 14 T frequency. The value of x- and y-axis intercepts are shown by the arrows.}
\end{figure}
\begin{figure}
\includegraphics[width=0.5\textwidth]{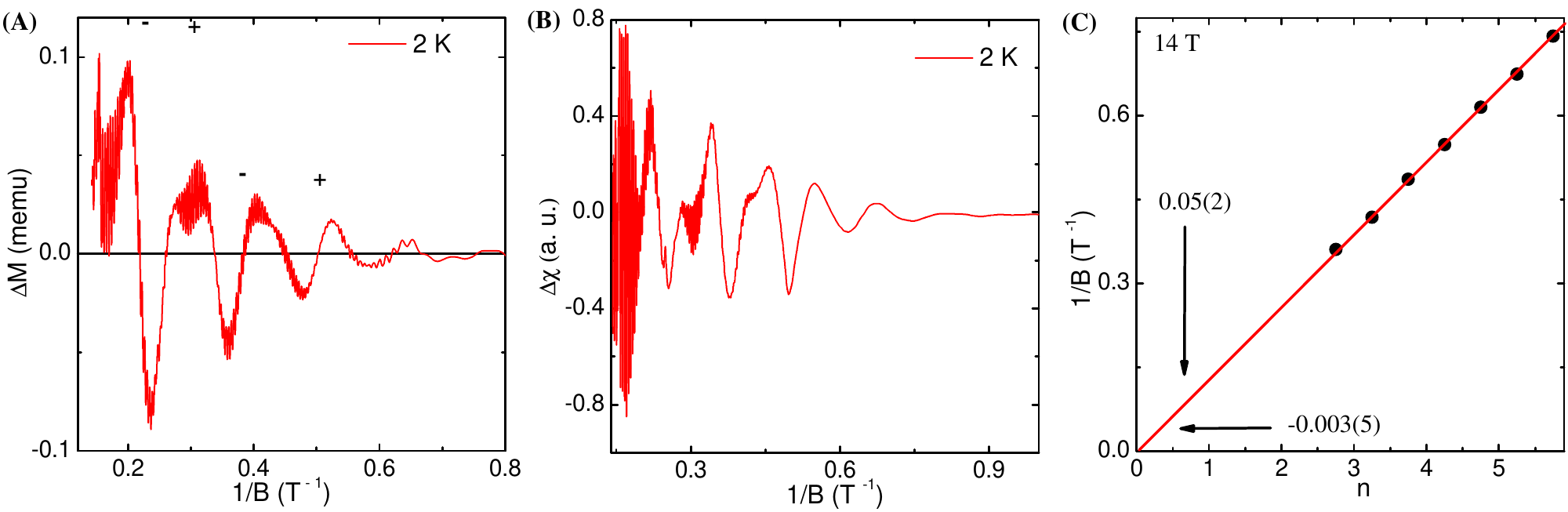}
\renewcommand{\figurename}{Fig.S}
\caption{\textit{(A)} The dHvA oscillation at 2 K, obtained after background subtraction. \textit{(B)} Oscillatory part of the magnetic susceptibility $\Delta$$\chi$ for the smaller frequency. \textit{(C)} The Landau index \textit{n} + 1/4 plot against 1/\textit{B} for the smaller frequency. The value of x- and y-axis intercepts are shown by the arrows.}
\end{figure}

\begin{figure}
\includegraphics[width=0.3\textwidth]{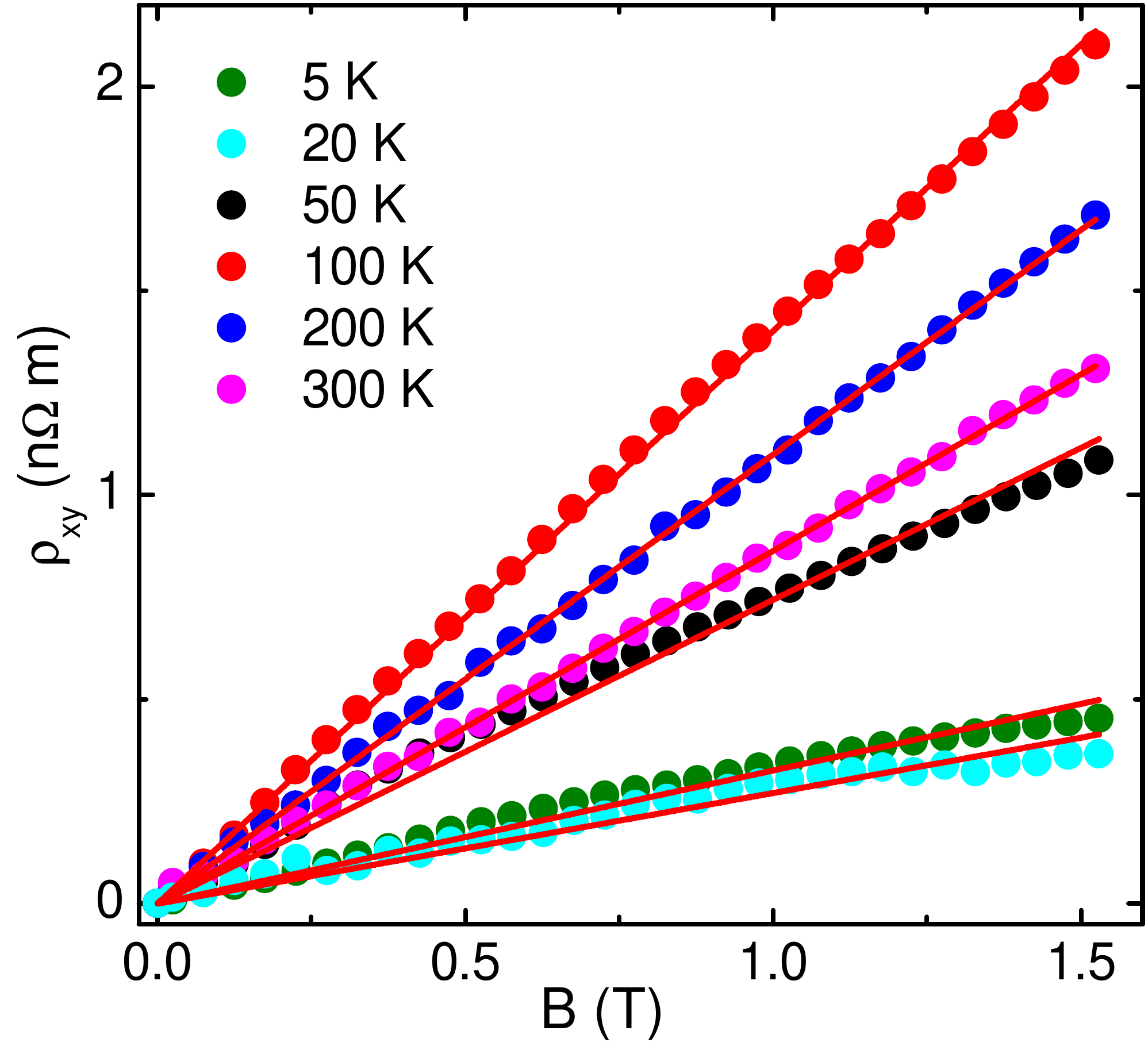}
\renewcommand{\figurename}{Fig.S}
\caption{Hall resistivity as a function of the magnetic field. The solid lines are classical two-band model fit to the experimental data.}
\end{figure}

\textbf{Two-band fitting of the Hall resistivity.} To calculate the classical Drude mobility of the charge carriers, the experimental Hall data are fitted (Fig. S8) using two-band model,
\begin{equation}
\rho_{xy} = \frac{B}{e} \frac{(n_{h}\mu_{h}^{2} - n_{e}\mu_{e}^{2})}{(n_{h}\mu_{h} + n_{e}\mu_{e})^{2}},
\end{equation}
in the low-field limit (45). $\mu_{h}$ ($\mu_{e}$) and \textit{n}$_{h}$ (\textit{n}$_{e}$) are the hole (electron) mobilities and densities, respectively. The calculated parameters are in agreement with those obtained from the quantum oscillation and listed in Table I of the main part of the article.\\

\textbf{Why it is important to study the other members of \textit{WHM} family ?} ZrSiS represents a large family of materials (\textit{WHM} with \textit{W} = Zr, Hf; \textit{H} = Si, Ge, Sn; \textit{M} = O, S, Se and Te) with identical crystal structure (PbFCl type). Recently, this family has been theoretically proposed as potential candidate for 2D topological insulator (19). However, subsequent ARPES measurement has revealed multiple bulk band crossings with linear dispersion over a wide energy range and a three-dimensional Dirac line node in ZrSiS (20). Topological nodal line fermions have been observed in very few systems, where the strength of spin-orbit coupling (SOC) plays a crucial role in protecting the line nodes (18). Therefore, with increasing atomic number, for example from Zr (Z=40) to Hf (Z=72), it is expected that the increasing SOC will substantially affect the topological protection of the non-trivial electronic band structure. Furthermore, as \textit{M} changes from O to Te, it has been seen that the particle-hole asymmetry increases and the interlayer binding energy decreases (19). While lower particle-hole asymmetry introduces a global band gap, weaker interlayer coupling enhances the mono-layer characteristic in electronic properties of the system. So it may experimentally possible to realize 2D topological insulating phase in other members of the family and require further investigations. In this context, the present work explicitly describes the different unusual electronic properties in ZrSiS and may pave the way to subsequent rigorous study on other members of this group.

\end{document}